\documentclass[letterpaper, 10 pt, conference]{ieeeconf}  
\IEEEoverridecommandlockouts
\makeatletter
\let\NAT@parse\undefined
\makeatother

\usepackage[dvipsnames]{xcolor}

\newcommand*\linkcolours{ForestGreen}

\usepackage{times}
\usepackage{graphicx}
\usepackage{amssymb}
\usepackage{gensymb}
\usepackage{amsmath}
\usepackage{breakurl}
\usepackage{fixltx2e}

\usepackage{url,hyperref}
\hypersetup{
colorlinks,
linkcolor=\linkcolours,
citecolor=\linkcolours,
filecolor=\linkcolours,
urlcolor=\linkcolours}

\usepackage{algorithm}
\usepackage{algorithmic}

\usepackage[labelfont={bf},font=small]{caption}
\usepackage[none]{hyphenat}

\usepackage{mathtools, cuted}

\usepackage[noadjust, nobreak]{cite}

\usepackage{tabularx}
\usepackage{amsmath}

\usepackage{float}

\usepackage{pifont}

\newcolumntype{Y}{>{\centering\arraybackslash}X}

\usepackage[]{placeins}

\usepackage{placeins}

\usepackage{tikz}

\usepackage[framemethod=tikz]{mdframed}

\usepackage{afterpage}

\usepackage{stfloats}

\usepackage{atbegshi}
\newcommand{\handlethispage}{}
\newcommand{\discardpagesfromhere}{\let\handlethispage\AtBeginShipoutDiscard}
\newcommand{\keeppagesfromhere}{\let\handlethispage\relax}
\AtBeginShipout{\handlethispage}

\usepackage{comment}

\title{\LARGE \bf
A Deep learning Approach to Generate Contrast-Enhanced Computerised Tomography Angiography without the Use of Intravenous Contrast Agents}

\author{Anirudh Chandrashekar$^{1*}$, Ashok Handa$^{1*}$, Natesh Shivakumar$^{1}$, Pierfrancesco Lapolla$^{1}$, \\ Vicente Grau$^{2**}$ and Regent Lee$^{1**}$
\thanks{$^{1}$ Nuffield Department of Surgical Sciences, University of Oxford}%
\thanks{$^{2}$ Department of Engineering Science, University of Oxford}%
\thanks{$^{*,**}$ Authors contributed equally.}%
}

\begin{document}

\maketitle
\thispagestyle{empty}
\pagestyle{empty}

\begin{abstract}
Contrast-enhanced computed tomography angiograms (CTAs) are widely used in cardiovascular imaging to obtain a non-invasive view of arterial structures. However, contrast agents are associated with complications at the injection site as well as renal toxicity leading to contrast-induced nephropathy (CIN) and renal failure. We hypothesised that the raw data acquired from a non-contrast CT contains sufficient information to differentiate blood and other soft tissue components. We utilised deep learning methods to define the subtleties between soft tissue components in order to simulate contrast enhanced CTAs without contrast agents. Twenty-six patients with paired non-contrast and CTA images were randomly selected from an approved clinical study. Non-contrast axial slices within the AAA from 10 patients (n = 100) were sampled for the underlying Hounsfield unit (HU) distribution at the lumen, intra-luminal thrombus and interface locations. Sampling of HUs in these regions revealed significant differences between all regions (p<0.001 for all comparisons), confirming the intrinsic differences in the radiomic signatures between these regions. To generate a large training dataset, paired axial slices from the training set (n=13)  were augmented  to produce a total of 23,551 2-D images. We trained a 2-D Cycle Generative Adversarial Network (cycleGAN) for this non-contrast to contrast (NC2C) transformation task. The accuracy of the cycleGAN output was assessed by comparison to the contrast image. This pipeline is able to differentiate between visually incoherent soft tissue regions in non-contrast CT images. The CTAs generated from the non-contrast images bear strong resemblance to the ground truth. Here we describe a novel application of Generative Adversarial Network for CT image processing. This is poised to disrupt clinical pathways requiring contrast enhanced CT imaging.

\end{abstract}

\section{INTRODUCTION}
Since its introduction in the 1970s, computed tomography (CT) has become widely used in medical imaging to obtain a comprehensive and non-invasive view of internal structures and has revolutionized diagnostic decision making. Especially in the field of  surgery,  CT imaging decreased the need for emergency procedures from 13\% to 5\% and has significantly decreased the need for exploratory surgical procedures \cite{Power2016}. Furthermore, its incorporation into clinical practice has optimized hospital workflow by decreasing the number of patients requiring inpatient care \cite{Rosen2000,Rosen2003}. In the NHS alone, 6 million CT scans were performed in 2018-2019 \cite{Baker2020}. The basic principle of CT technology involves transmitting ionizing radiation, or x-rays, through a region-of-interest (ROI). The transmitted rays are then incident on an electronic detector to create a ‘cut’ through the object. Both the radiation source and the detector rotate around the object to obtain multiple ‘slices or ‘cuts’. These projections are then used to reconstruct a 3-D representation of the ROI. Recent advancements have allowed for faster acquisition times, less intrinsic movement artefact and the ability to capture a greater area at a higher resolution during a single acquisition period \cite{Foley,Sun2012}.

As the x-rays pass through the patient, the rays are attenuated depending on the density of tissue it passes through. The variations in physical density of different objects translates to differences in attenuation and subsequent radio- densities  (measured  in  Hounsfield  Units,  HU)  on a CT scan\cite{Foley}. The higher the attenuation, the brighter CT image (ex. Bone and calcification) and the lower the attenuation, the darker the CT image (ex. Air). Therefore, the intrinsic contrast of the image is  generated based  on the differences in attenuation between adjacent tissues \cite{Sun2012}. 

\textbf{Figure 1} gives an example of  an  axial  CT  slice  through the abdomen. Bony spine (orange arrow) is distinctly more radio-opaque compared to psoas muscle (blue arrow) or the abdominal aorta (red arrow). However, the densities between the abdominal aorta and psoas muscles appear similar here. 

\begin{figure}[tbh!]
\centering
\includegraphics[width=0.98\columnwidth]{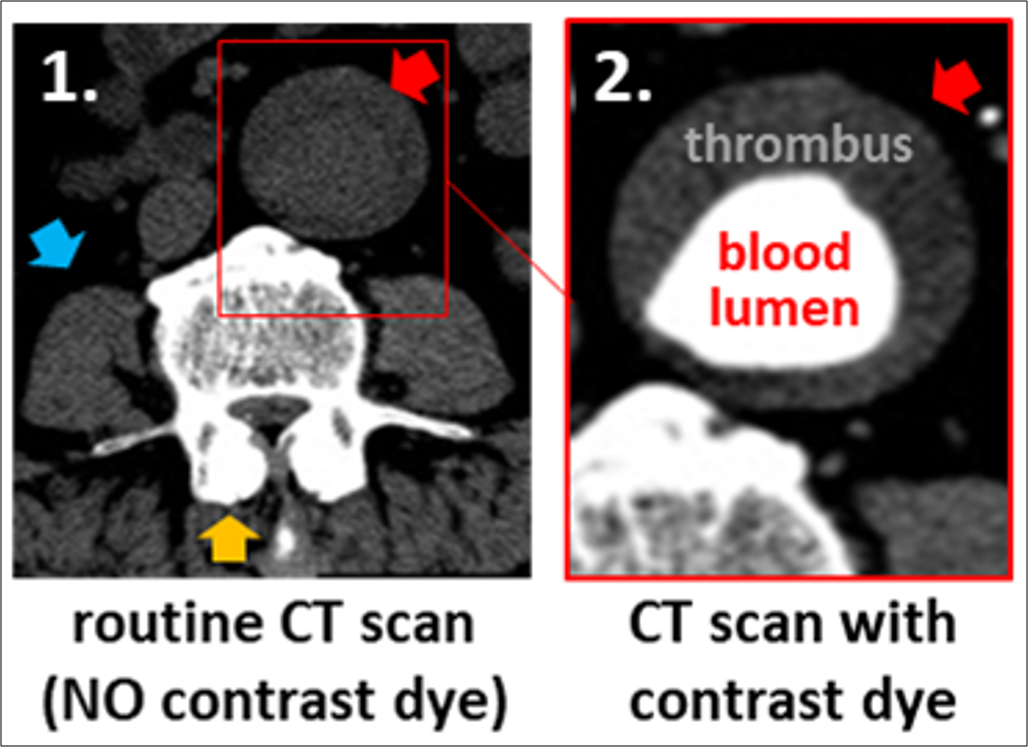}
\caption{Axial slice from a Computed Tomography (CT) scan with and without the use of an intravenous contrast agent, which enhances visualization of the vasculature and allows for diagnosis of vascular pathology.}
\label{Figure 1}
\end{figure}

Where treatment of an artery is being considered, a detailed view of the arterial anatomy is required. In the example of abdominal aortic aneurysms (AAA, abnormal ballooning of the abdominal aorta), an intra-luminal thrombus (ILT) adherent to the aortic wall within the enlarging aneurysmal sac is present in 95\% of cases \cite{Aggarwal2011a}. Given the similarities in density between the blood lumen, ILT and the complex blood-thrombus interface, these regions cannot be readily distinguished using a conventional CT image.Clear visualization of these regions can only be achieved by the injection of an intravenous (IV) contrast agent. The primary purpose of IV contrast is to enhance luminal density and increase both the attenuation and intrinsic contrast between the vascular tree and surrounding soft tissues. This optimizes the visualization of the vasculature and generates  a CT angiogram (CTA) \cite{Foley,Sun2012}. 

Although CTAs may provide unique insight into the structure of the vascular tree, it is associated with several disadvantages \cite{Sun2012,Hinson2017}. CTAs are contraindicated in patients with iodine allergies as most agents are iodine-based. Furthermore, administration of contrast agents requires needle insertion. This causes additional discomfort  and  has  been  associated with complications including inadvertent arterial puncture by needle, and contrast leak from veins causing skin irritation/damage \cite{Hinson2017}. 
Additionally, contrast agents are nephrotoxic and have up to 12\% incidence of acute kidney injury (contrast-induced nephropathy) following use \cite{Hinson2017}. This is especially a problem within the elderly population, who either have decreasing baseline renal function or concomitant chronic kidney disease. In these high-risk patients, there is a recognised risk of complete kidney failure, which may lead to renal dialysis.

We hypothesise that the raw data acquired from a non-contrast CT contains sufficient information to differentiate blood and other soft tissue components. Blood, thrombus, and artery wall are made up of different components: blood is predominantly fluid, with red/white blood cells; thrombus is predominantly fibrinous and collagenous, with red cells/platelets; artery wall predominantly contains smooth muscle cells with collagen. These individual components vary in physical density which should reflect in different (albeit subtle) HUs on a CT scan. We further hypothesise that using deep learning approaches, the subtleties between the various components of the soft tissue can be defined and amplified to enable simulation     of contrast enhanced CT images without the need to inject contrast agents.

\section{Methods}
\subsection{\textit{Obtaining CT images from a clinical cohort}}
Computerised Tomographic scans of the chest and abdomen were acquired through the Oxford Abdominal Aortic Aneurysm (OxAAA) study. The study received full regulatory and ethics approval from both Oxford University and Oxford University Hospitals (OUH) National Health Services (NHS) Foundation Trust (Ethics Ref 13/SC/0250). As part of the routine pre-operative assessment for aortic aneurysmal disease, a non-contrast CT of the chest/abdomen/pelvis and an arterial phase CT angiogram (CTA) was performed. CTA images were obtained following contrast injection in helical mode with a pre-defined slice thickness of 1.25 mm. On the other hand, non-contrast CT images were obtained with a pre-defined slice thickness of 2.5 mm. Paired contrast and non-contrast CT images were anonymised within the OUH PACS system before being downloaded onto the secure study drive. Twenty-six patients with paired non-contrast and CTA images of the abdominal region were randomly selected.

\subsection{\textit{Manual Segmentation of CT Images (Training Dataset)}}
Thirteen (of the 26) cases were randomly selected as the training dataset. Manual segmentation of the aortic inner lumen (in contrast enhanced CTAs) and aortic outer lumen (for both non-contrast and contrast enhanced CTAs) were performed using the open source ITK snap software \cite{Yushkevich2006}. Segmentation of the aorta was performed from the aortic root to the iliac bifurcation.   

\subsection{\textit{Reorientation of Contrast-Enhanced Scans}}
To account for voluntary and involuntary movement by the patient between scans, it was necessary to re-orientate the contrast-enhanced images obtained and corresponding segmentation masks to the non-contrast image plane. Registration of the axial slices was performed using the Image Orientation Patient (X, Y, Z coordinates) and Image Orientation Patient (patient’s relative rotation) attributes found within the DICOM header. Minimal variation in the orientation of the bowel/air bubbles was observed between the contrast and non-contrast slices. Appropriate registration of 100 randomly selected axial images was confirmed by 2 blinded reviewers (NS and PL) slices prior to subsequent analysis.

\subsection{\textit{Hounsfield Unit Sampling}}
In order to investigate the regional differences, non-contrast axial slices within  the  aneurysmal  region,  10 consecutive axial slices were taking from 10 patients,  (total = 100 slices)  were  sampled  for the underlying Hounsfield unit (HU) distribution at the lumen, intra-luminal thrombus and interface  locations (\textbf{Figure 2}). These visually indistinct regions on the non-contrast CT slice were identified from their paired contrast counterpart. The reoriented binary masks were used to establish the boundary locations in the non-contrast image. To account  for slight inaccuracies in the image  reorientation  process and minimize sampling errors, the thrombus (blue) and lumen (yellow) areas were reduced by 20\% and the zone between the two regions was demarcated as the interface (red). This delineation is clearly indicated in \textbf{Figure 2}.

\begin{figure}[tbh!]
\centering
\includegraphics[width=1\columnwidth]{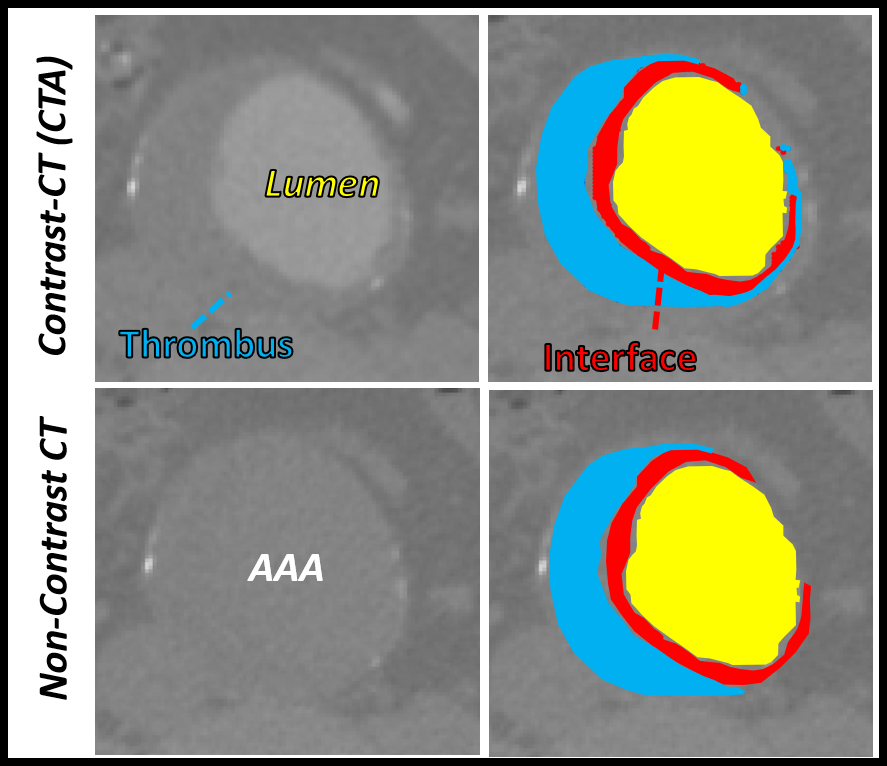}
\caption{Paired Non-Contrast and CTA images with clearly demarcated boundary regions used for regional Hounsfield unit sampling. Regions highlighted included the inner lumen (yellow), intra-luminal thrombus (blue) and the blood-thrombus interface (red).}
\label{Figure 2}
\end{figure}

One-way ANOVA was used to compare the average HU intensity within each region (in individual patients) both across and  within  each slice for each patient. The negative control consisted   of concentric sampling within the lumen region. This was used as blood on a macroscopic level is randomly distributed and relatively homogeneous and should produce similar HU intensities.

\subsection{\textit{Image Pre-Processing}}
Of the 26 patients, 13 patients were randomly allocated   to the training (n\textsubscript{train} = 13) cohort. Following segmentation, the original pre-aligned CT images and their corresponding image masks of patients in the training cohort were augmented using divergence transformations. These divergence transformations employ non-linear warping techniques to each axial slice, which manipulate the aorta in certain predefined locations. In this instance, both congruent and divergent local transformations were utilized to diversify the training dataset. These augmentation methods have been previously validated. Therefore, each patient’s scan in the training cohort was augmented in a ratio of 10:1 to obtain  a total of 23,551 2-D images (axial).

\subsection{\textit{Cycle-GAN Architecture}} 
In this study, we utilized a cycle-GAN, which is a variation of the popular generative adversarial network (GAN). These networks are a class of deep learning architecture whereby two neural networks train simultaneously, with one network focused on data generation (generator) and the other network focused on data discrimination (discriminator). The two neural networks ‘compete’ against each other, learning the statistical distribution of the training data, which in turn allows to generate new examples from the same distribution. GANs have been applied to generate imaginary portraits, landscapes, and artworks based on real examples \cite{Zhu2017a}. Many variations of the original GAN concept have been developed, including conditional GANs (cGANs), which can learn the transformation between two paired distributions, including the transformation between images using the pixel to pixel (Pix2Pix) approach \cite{Isola2016}. The primary benefit of cylceGANs is that it can learn transformations from two distributions without the need for direct pairings between specific samples \cite{Zhu2017a}.

The generator and discriminator components in the NonContrast-to-Contrast (NC2C) model architecture (\textbf{Figure 3}) were explicitly defined as least-squares GAN and a 70 x 70 PatchGAN, respectively. The former incorporates an additional least-squares loss function for the discriminator, which in turn, improves the training of the generative model. On the other hand, the discriminator goes through the image pairs, in 70 x 70 patches, and is trained to classify whether the image under question is “real” or “fake”.

\begin{figure}[tbh!]
\centering
\includegraphics[width=1\columnwidth]{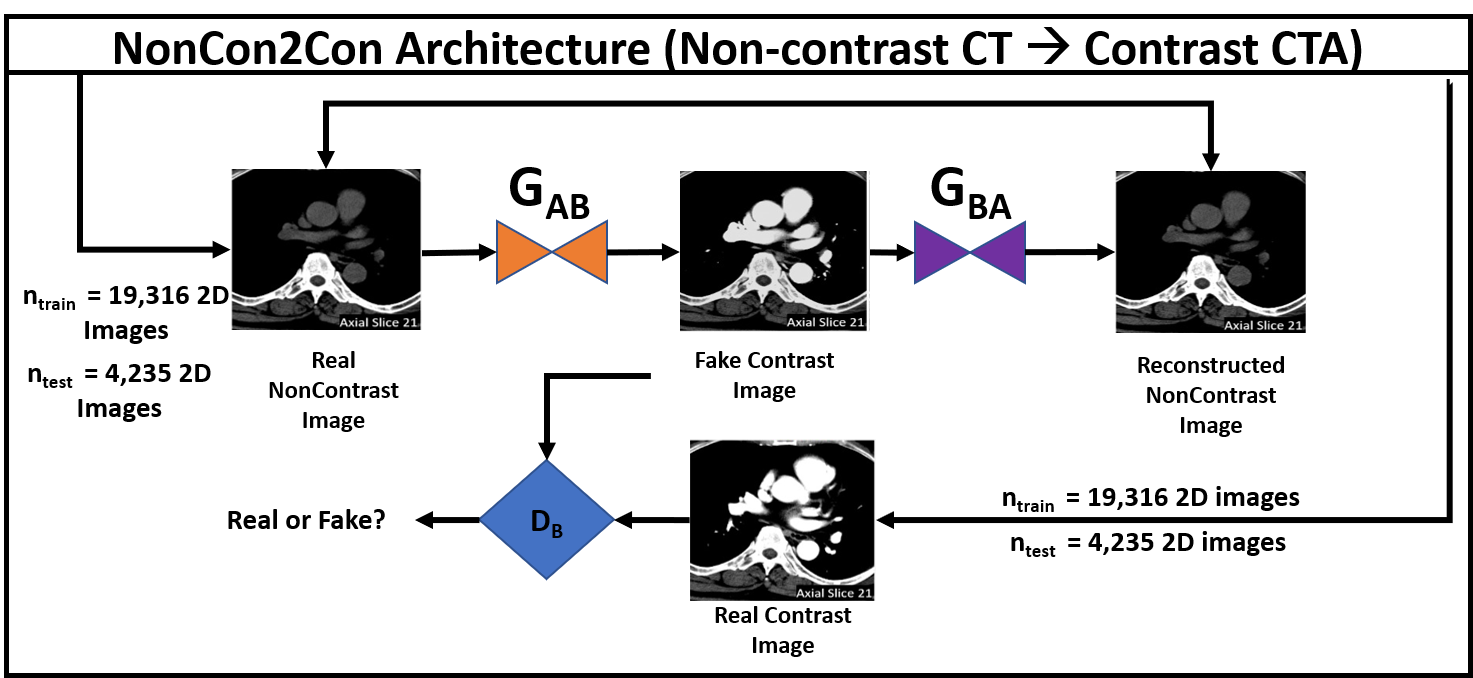}
\caption{NonCon2Con (NC2C) Cycle-GAN architecture used for the transformation of non-contrast axial CT images to contrast CTA images.}
\label{Figure 3}
\end{figure}

\subsection{\textit{Model Training and Evaluation}}
The NC2C-cycleGAN (\textbf{Figure 3}) was trained with a learning rate of 2.0 * 10\textsuperscript{-4} for 200 epochs on 256 x 256 images centred around the aorta. Four networks (2 generators + 2 discriminators) were trained simultaneously and various loss functions were evaluated at each iteration to document model training. In addition to the loss metrics inherent to the networks, both an identity mapping and a cycle consistency loss functions were included to ensure appropriate style transfer and regularization of the generator to allow for image translation, respectively. The accuracy of the cycleGAN output in generating the lumen and thrombus interface was assessed by comparison to the contrast image, which serves as a gold standard.

\begin{figure*}[tbh]
\centering
\includegraphics[width=0.99\textwidth]{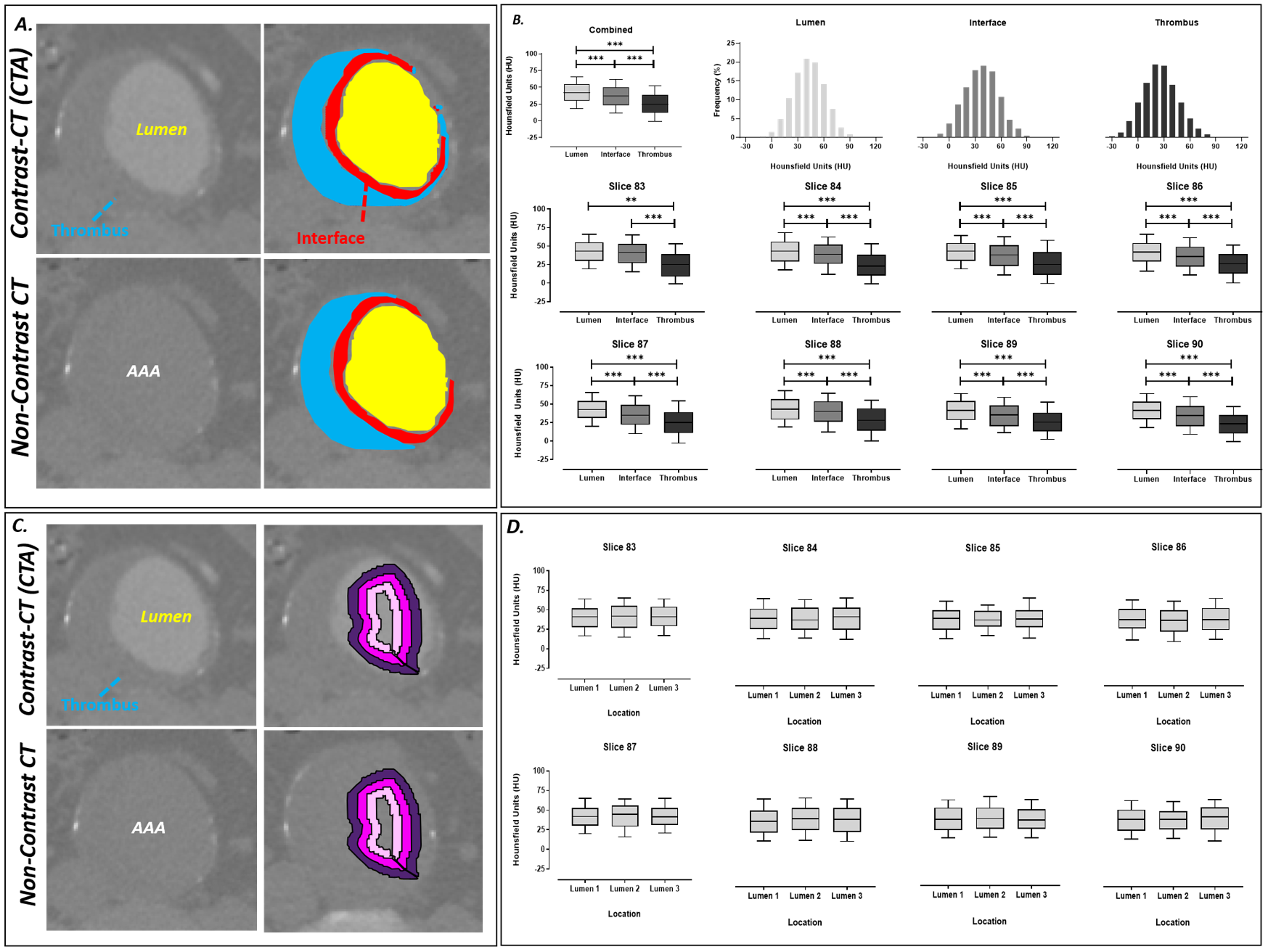}
\caption{ Axial slices from both the Contrast and Non-contrast CT scans. \textbf{A.} Demarcated regions display the thrombus (blue), lumen (yellow) and the interface between the regions (red). \textbf{B.}. Hounsfield unit sampling of the lumen, interface and thrombus with histograms displaying the frequency of HUs within each region. Analysis at each axial slice was performed and also showed differences in HU intensity between the lumen, interface and thrombus regions. \textbf{C.} Concentric sampling within the lumen as demarcated by the pink, magenta and purple, was used as the negative control for this experiment \textbf{D.} Hounsfield unit sampling of the lumen at multiple locations indicated minimal difference in HU intensity.}
\label{Figure 4}
\end{figure*}

\section{Results}
Average HU intensity in the non-contrast images was significantly different between all three regions (Lumen vs. Thrombus, Lumen vs. Interface and Interface vs. Thrombus) for all patients assessed. When assessing on a slice-by-slice basis, the average HU intensity of the thrombus was significantly different from that of the lumen 94\% of the time. Histograms corresponding to the HU frequencies for each region had considerable overlap, however a shift in frequency distribution is also apparent (\textbf{Figure 4B}). \textbf{Figure 4B} further displays the differences in HU intensities for 8 axial slices obtained from one of the ten patients analyzed. The average HU intensity for the thrombus was significantly lower when compared to that of the lumen for all axial slices sampled. Furthermore, the interface was also significantly different from the other two regions, indicating a gradual change in pixel intensity from the center lumen to the peripheral thrombus. No significant differences in HU intensity were noted following concentric sampling  of the aortic blood flow lumen (\textbf{Figure 4C,D}). 

\textbf{Figure 5} illustrates the loss functions during the training of the NC2C-cycle GAN. The inherent generator and discriminator loss functions converge during model training. Additionally, the identity mapping and cycle consistency loss functions for both directions (1. Contrast-to-Non-Contrast, 2. Non-Contrast-to-Contrast) plateau within the first 50 epochs. Model performance was evaluated on the testing (n\textsubscript{test} =13) cohort. This generative model is able to  simulate the aortic lumen throughout the length of the aorta and differentiate between the lumen and intra-luminal thrombus of aneurysmal sections with strong resemblance to the ground truth. In addition to the aorta, this algorithm is able to transform other structures including the small mesenteric arteries (orange arrows), the pulmonary arteries (yellow arrow) and the kidneys (blue arow) as seen in \textbf{Figure 6}.

\begin{figure}[tbh!]
\centering
\includegraphics[width=1\columnwidth]{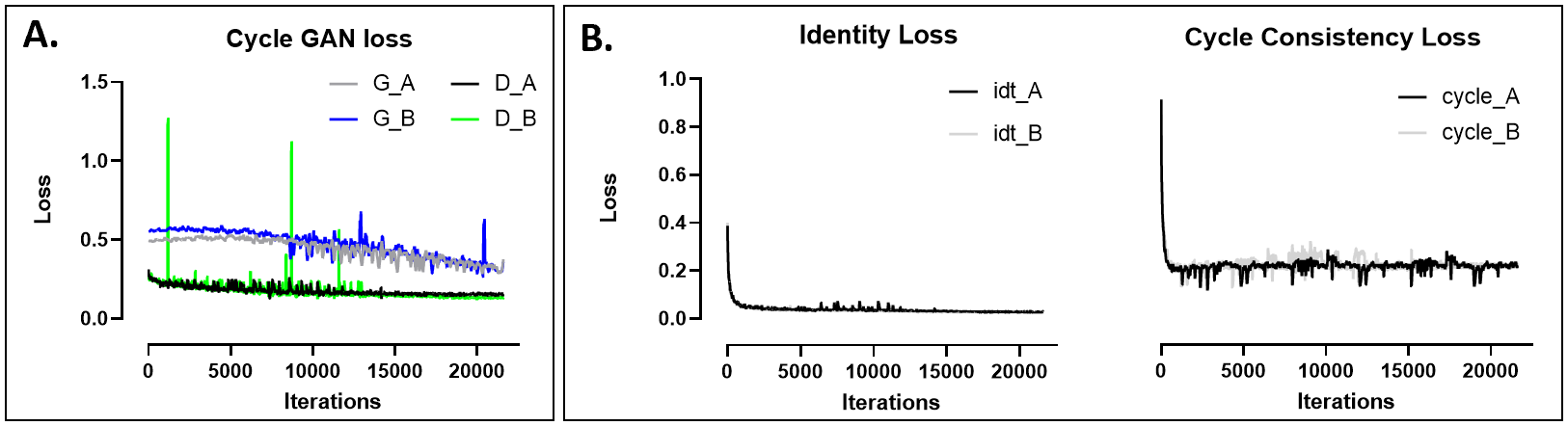}
\caption{\textbf{A.} Loss functions associated with the NC2C cycle GAN architecture. \textbf{B.} Identity mapping and cycle consistency loss functions were calculated iteratively during model training to ensure style transfer and regularization of the generator.}
\label{Figure 6}
\end{figure}

\begin{figure*}[tbh]
\centering
\includegraphics[width=1\textwidth]{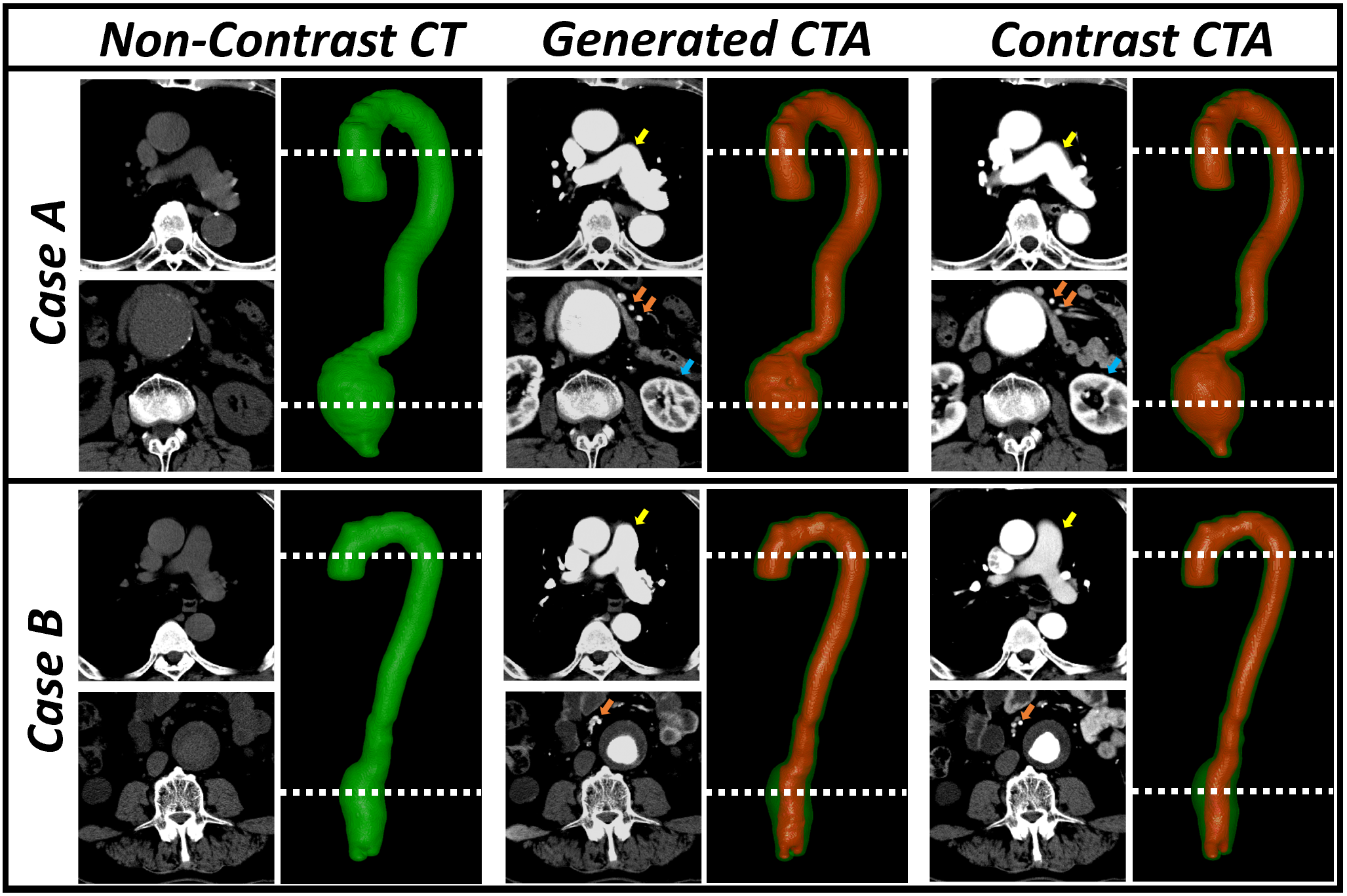}
\caption{ NC2C-generated CTA images alongside their ground truth CTA and non-contrast axial CT image for two patients \textbf{(Case A,B)}. 3D-Volume reconstructions of the aorta highlight the ability of this GAN in isolating the aortic inner lumen and differentiating it from the pathological thrombus within the aneurysm region.}
\label{Figure 6}
\end{figure*}

\section{Discussion}

In the realm of medical image synthesis, traditional image transformation tasks have relied on patch-based regressions. These methods are able to sample a patch of an image or volume from one modality and predict its intensity in the target modality \cite{Jog2013, Torrado-Carvajal2016}. In addition to patch-based regression models, sparse representation \cite{Huang2017} and atlas-based models \cite{Miller1993} have been used. The latter utilizes paired  image   atlases from the source- and  target  image  modalities. Here, the atlas-to-image transformation is calculated for the source-modality and a similar transformation is performed for the target-modality atlas \cite{Miller1993,Roy2013,Burgos2014}

In  recent  years,  deep  learning  (DL)  methods  have been applied extensively in this domain and has achieved  promising results. Dong et al. proposed a convolutional neural network approach to increase the resolution of medical images \cite{Dong2014}. Additionally, GAN-based cross-modal image synthesis methods have gained significant traction and achieved success \cite{Nie2016}. Here, these key methods rely on an adversarial training platform between   a   generator   and a  discriminator.  The  fundamental  task  of  the  generator  is to  produce  images  that  are  realistic  and  ultimately  fool the discriminator, which is simultaneously trained to differentiate  real  and  generated  images  \cite{Goodfellow2014}.  Variations  of the GAN architecture include the Pix2Pix \cite{Isola2016}, and conditional \cite{Mirza2014} and cycle -GAN networks \cite{Zhu2017a}. As its name suggests, the Pix2pix network uses paired data to learn the appropriate pixel-to-pixel image transformations. This network strives to maintain pixel-to-pixel similarity between the real and the generated images \cite{Isola2016}. On the other hand, the CycleGAN attempts to capture structural information by learning the translation mapping between unpaired images \cite{Zhu2017a}. GAN-based methods have shown great promise in synthesizing various types of  medical  images  including  CT (from MR  images \cite{Nie2016},  low-dose  to  routine-dose \cite{Yang2017}), retinal \cite{Costa2017,Costa2018}, MR, and ultrasound images \cite{Tom2017}.  This study attempts, for the first time, the possibility to transform non-contrast CT images into contrast-enhanced CT Angiograms without the need the for intravenous (IV) contrast.

Currently, CT angiography relies on IV contrast injection to enhance the intrinsic contrast between the vascular tree and surrounding tissues. Therefore, this method is able to provide a unique view of the patient’s vasculature, which is essential for diagnosis and surgical planning. For example, in the case of AAA disease, CTAs are the current gold-standard to visualize the intra-luminal thrombus (ILT) within the enlarging aneurysmal sac. The orientation, location and extent of the ILT as well as the complex blood-ILT interface is vital information prior to surgical intervention \cite{Aggarwal2011a}. Although CT angiography may provide unique insight into aneurysm morphology and the structure of the vascular tree, it is not without its disadvantages, as discussed above. 

Additionally, this method to transform non-contrast CT to CTA images can be used to incorporate historic non-contrast CT scans for research purposes. A key priority for AAA research is to discovery novel indices of AAA growth prediction. Currently, this is documented by measuring the maximum antero-posterior aortic diameter of the abdominal aorta during serial duplex ultrasound scans \cite{Schlosser2008}. However, this uses a 1-dimensional measurement that is susceptible to inter-observer error to characterize a diverse and complex 3-dimensional growing  process  \cite{Kitagawa2013}.  There  is emerging evidence that patient-specific geometric and volumetric measurements more readily influence AAA growth \cite{Shum2011}. As small AAAs enlarge, a variety of geometrical changes have been observed. Many of these changes result  in a unique non-uniform distribution of wall stress and have been hypothesized to either favour AAA growth deceleration or increase rupture risk \cite{Shum2011,Martufi2013}. Isolating and deciphering these changes will allow us to predict AAA growth and progression in each patient. Current work by our group has focused on generating a DL-algorithm for the automatic aortic segmentation of the aortic volume alongside feature extraction \cite{Chandrashekar2020}. Integration of this image synthesis model with our volume extraction methods will allow us to utilize historic non-contrast CT scans across multiple centers for this complex geometric and morphological analysis.

This image synthesis pipeline can be extended to cover other anatomical structure (veins, solid organs, etc). Our preliminary work has shown that it is capable of generating reliable CTAs of the large arteries; however, there is evidence that this model is able to account for small vessels that branch from the aorta (ex. renal arteries, vertebral arteries, etc. and can be extended to other structures (ex. venogram, solid organ contrast CTs).

\section{Conclusions}

This study describes, for the first time, the ability to differentiate between visually incoherent soft tissue regions in non-contrast CT images using deep learning methods. Ultimately, refinement of this methodology may negate the use of intravenous contrast and prevent related complications.


\section{Acknowledgements}

We acknowledge the support from the following: Medical Sciences Division, University of Oxford Medical Research Fund; John Fell Fund, University of Oxford; Academy of Medical Sciences Starter Grant to RL (AMS-SGL013 \textbackslash1015; Oxford University Clarendon Scholarship to AC. The methods described in this manuscript is subject to a patent filing (UK priority filing, P276235GB).


\bibliographystyle{ieeetr}
\bibliography{NC2C.bib}

\begin{thebibliography}{10}

\bibitem{Power2016}
S.~P. Power, F.~Moloney, M.~Twomey, K.~James, O.~J. O'Connor, and M.~M. Maher,
  ``{Computed tomography and patient risk: Facts, perceptions and
  uncertainties.},'' {\em World journal of radiology}, vol.~8, pp.~902--915,
  dec 2016.

\bibitem{Rosen2000}
M.~P. Rosen, D.~Z. Sands, H.~E. Longmaid, K.~F. Reynolds, M.~Wagner, and
  V.~Raptopoulos, ``{Impact of abdominal CT on the management of patients
  presenting to the emergency department with acute abdominal pain.},'' {\em
  AJR. American journal of roentgenology}, vol.~174, pp.~1391--6, may 2000.

\bibitem{Rosen2003}
M.~P. Rosen, B.~Siewert, D.~Z. Sands, R.~Bromberg, J.~Edlow, and
  V.~Raptopoulos, ``{Value of abdominal CT in the emergency department for
  patients with abdominal pain.},'' {\em European radiology}, vol.~13,
  pp.~418--24, feb 2003.

\bibitem{Baker2020}
C.~Baker, ``{NHS Key Statistics},'' {\em House of Commons Library}, vol.~7281,
  2020.

\bibitem{Foley}
W.~D. Foley and M.~Karcaaltincaba, ``{Computed tomography angiography:
  principles and clinical applications.},'' {\em Journal of computer assisted
  tomography}, vol.~27 Suppl 1, pp.~S23--30.

\bibitem{Sun2012}
Z.~Sun, G.~H. Choo, and K.~H. Ng, ``{Coronary CT angiography: current status
  and continuing challenges.},'' {\em The British journal of radiology},
  vol.~85, pp.~495--510, may 2012.

\bibitem{Aggarwal2011a}
S.~Aggarwal, A.~Qamar, V.~Sharma, and A.~Sharma, ``{Abdominal aortic aneurysm:
  A comprehensive review.},'' {\em Experimental and clinical cardiology},
  vol.~16, no.~1, pp.~11--5, 2011.

\bibitem{Hinson2017}
J.~S. Hinson, M.~R. Ehmann, D.~M. Fine, E.~K. Fishman, M.~F. Toerper, R.~E.
  Rothman, and E.~Y. Klein, ``{Risk of Acute Kidney Injury After Intravenous
  Contrast Media Administration.},'' {\em Annals of emergency medicine},
  vol.~69, pp.~577--586.e4, may 2017.

\bibitem{Yushkevich2006}
P.~A. Yushkevich, J.~Piven, H.~C. Hazlett, R.~G. Smith, S.~Ho, J.~C. Gee, and
  G.~Gerig, ``{User-guided 3D active contour segmentation of anatomical
  structures: significantly improved efficiency and reliability.},'' {\em
  NeuroImage}, vol.~31, pp.~1116--28, jul 2006.

\bibitem{Zhu2017a}
J.-Y. Zhu, T.~Park, P.~Isola, and A.~A. Efros, ``{Unpaired Image-to-Image
  Translation using Cycle-Consistent Adversarial Networks},'' mar 2017.

\bibitem{Isola2016}
P.~Isola, J.-Y. Zhu, T.~Zhou, and A.~A. Efros, ``{Image-to-Image Translation
  with Conditional Adversarial Networks},'' nov 2016.

\bibitem{Jog2013}
A.~Jog, S.~Roy, A.~Carass, and J.~L. Prince, ``{MAGNETIC RESONANCE IMAGE
  SYNTHESIS THROUGH PATCH REGRESSION.},'' {\em Proceedings. IEEE International
  Symposium on Biomedical Imaging}, vol.~2013, pp.~350--353, dec 2013.

\bibitem{Torrado-Carvajal2016}
A.~Torrado-Carvajal, J.~L. Herraiz, E.~Alcain, A.~S. Montemayor,
  L.~Garcia-Ca{\~{n}}amaque, J.~A. Hernandez-Tamames, Y.~Rozenholc, and
  N.~Malpica, ``{Fast Patch-Based Pseudo-CT Synthesis from T1-Weighted MR
  Images for PET/MR Attenuation Correction in Brain Studies.},'' {\em Journal
  of nuclear medicine : official publication, Society of Nuclear Medicine},
  vol.~57, pp.~136--43, jan 2016.

\bibitem{Huang2017}
Y.~Huang, L.~Shao, and A.~F. Frangi, ``{Simultaneous Super-Resolution and
  Cross-Modality Synthesis of 3D Medical Images using Weakly-Supervised Joint
  Convolutional Sparse Coding},'' may 2017.

\bibitem{Miller1993}
M.~I. Miller, G.~E. Christensen, Y.~Amit, and U.~Grenander, ``{Mathematical
  textbook of deformable neuroanatomies.},'' {\em Proceedings of the National
  Academy of Sciences of the United States of America}, vol.~90, pp.~11944--8,
  dec 1993.

\bibitem{Roy2013}
S.~Roy, A.~Carass, and J.~L. Prince, ``{Magnetic Resonance Image Example-Based
  Contrast Synthesis.},'' {\em IEEE transactions on medical imaging}, vol.~32,
  pp.~2348--63, dec 2013.

\bibitem{Burgos2014}
N.~Burgos, M.~J. Cardoso, K.~Thielemans, M.~Modat, S.~Pedemonte, J.~Dickson,
  A.~Barnes, R.~Ahmed, C.~J. Mahoney, J.~M. Schott, J.~S. Duncan, D.~Atkinson,
  S.~R. Arridge, B.~F. Hutton, and S.~Ourselin, ``{Attenuation correction
  synthesis for hybrid PET-MR scanners: application to brain studies.},'' {\em
  IEEE transactions on medical imaging}, vol.~33, pp.~2332--41, dec 2014.

\bibitem{Dong2014}
C.~Dong, C.~C. Loy, K.~He, and X.~Tang, ``{Image Super-Resolution Using Deep
  Convolutional Networks},'' dec 2014.

\bibitem{Nie2016}
D.~Nie, R.~Trullo, C.~Petitjean, S.~Ruan, and D.~Shen, ``{Medical Image
  Synthesis with Context-Aware Generative Adversarial Networks},'' dec 2016.

\bibitem{Goodfellow2014}
I.~J. Goodfellow, J.~Pouget-Abadie, M.~Mirza, B.~Xu, D.~Warde-Farley, S.~Ozair,
  A.~Courville, and Y.~Bengio, ``{Generative Adversarial Networks},'' jun 2014.

\bibitem{Mirza2014}
M.~Mirza and S.~Osindero, ``{Conditional Generative Adversarial Nets},'' nov
  2014.

\bibitem{Yang2017}
Q.~Yang, P.~Yan, Y.~Zhang, H.~Yu, Y.~Shi, X.~Mou, M.~K. Kalra, and G.~Wang,
  ``{Low Dose CT Image Denoising Using a Generative Adversarial Network with
  Wasserstein Distance and Perceptual Loss},'' aug 2017.

\bibitem{Costa2017}
P.~Costa, A.~Galdran, M.~I. Meyer, M.~D. Abr{\`{a}}moff, M.~Niemeijer, A.~M.
  Mendon{\c{c}}a, and A.~Campilho, ``{Towards Adversarial Retinal Image
  Synthesis},'' jan 2017.

\bibitem{Costa2018}
P.~Costa, A.~Galdran, M.~I. Meyer, M.~Niemeijer, M.~Abramoff, A.~M. Mendonca,
  and A.~Campilho, ``{End-to-End Adversarial Retinal Image Synthesis.},'' {\em
  IEEE transactions on medical imaging}, vol.~37, no.~3, pp.~781--791, 2018.

\bibitem{Tom2017}
F.~Tom and D.~Sheet, ``{Simulating Patho-realistic Ultrasound Images using Deep
  Generative Networks with Adversarial Learning},'' dec 2017.

\bibitem{Schlosser2008}
F.~J.~V. Schl{\"{o}}sser, M.~J.~D. Tangelder, H.~J.~M. Verhagen, G.~J. M.~G.
  van~der Heijden, B.~E. Muhs, Y.~van~der Graaf, F.~L. Moll, and S.~study
  group, ``{Growth predictors and prognosis of small abdominal aortic
  aneurysms.},'' {\em Journal of vascular surgery}, vol.~47, pp.~1127--33, jun
  2008.

\bibitem{Kitagawa2013}
A.~Kitagawa, T.~M. Mastracci, R.~von Allmen, and J.~T. Powell, ``{The role of
  diameter versus volume as the best prognostic measurement of abdominal aortic
  aneurysms.},'' {\em Journal of vascular surgery}, vol.~58, pp.~258--65, jul
  2013.

\bibitem{Shum2011}
J.~Shum, G.~Martufi, E.~{Di Martino}, C.~B. Washington, J.~Grisafi, S.~C.
  Muluk, and E.~A. Finol, ``{Quantitative assessment of abdominal aortic
  aneurysm geometry.},'' {\em Annals of biomedical engineering}, vol.~39,
  pp.~277--86, jan 2011.

\bibitem{Martufi2013}
G.~Martufi, M.~Auer, J.~Roy, J.~Swedenborg, N.~Sakalihasan, G.~Panuccio, and
  T.~C. Gasser, ``{Multidimensional growth measurements of abdominal aortic
  aneurysms},'' {\em Journal of Vascular Surgery}, vol.~58, pp.~748--755, sep
  2013.

\bibitem{Chandrashekar2020}
A.~Chandrashekar, A.~Handa, N.~Shivakumar, P.~Lapolla, V.~Grau, and R.~Lee,
  ``{A Deep Learning Approach to Automate High-Resolution Blood Vessel
  Reconstruction on Computerized Tomography Images With or Without the Use of
  Contrast Agent},'' feb 2020.

\end{thebibliography}
\clearpage

\end{document}